\documentclass[
pra,aps,showpacs,nofootinbib]{revtex4}
\usepackage{epsfig}
\begin{document}

\newcommand{\JMP}{{\it J. Math. Phys}~}
\newcommand{\PRL}{{\it Phys. Rev. Lett.}~}
\newcommand{\PL}{{\it Phys. Lett.}~}
\newcommand{\PR}{{\it Phys. Rev.}~}
\newcommand{\PS}{{\it Phys. Scr.}~}
\newcommand{\JOB}{{\it J. Opt. B: Quantum Semiclass. Opt.} B~}

\title{
\vspace*{6cm} Dynamical Casimir Effect in a one-dimensional
uniformly contracting cavity}
\author{A.M. Fedotov\thanks{E-mail: fedotov@cea.ru},
Yu.E. Lozovik$^\dag$\thanks{Email: lozovik@isan.troitsk.ru}, N.B.
Narozhny\thanks{E-mail: narozhny@theor.mephi.ru}, and A.N.
Petrosyan\thanks{Email: petrossian777@yandex.ru}}
\address{Moscow Engineering Physics Institute,
115409 Moscow, Russia}
\address{$^\dag$Institute of Spectroscopy of Russian Academy of Science,
Troitsk, 142190 Moscow region, Russia}

\begin{abstract}
We consider particle creation (the Dynamical Casimir effect) in a
uniformly contracting ideal one-dimensional cavity
non-perturbatively. The exact expression for the energy spectrum of
created particles is obtained and its dependence on parameters of
the problem is discussed. Unexpectedly, the number of created
particles depends on the duration of the cavity contracting
non-monotonously. This is explained by quantum interference of the
events of particle creation which are taking place only at the
moments of acceleration and deceleration of a boundary, while stable
particle states exist (and thus no particles are created) at the
time of contracting.
\end{abstract}


\maketitle

\section{Introduction}

During the last two decades the dynamical Casimir effect (DCE), the
effect of photon creation in an empty nonstationary cavity
\cite{Moore}, attracted considerable attention in literature. Being
a firm prediction of quantum field theory, the DCE has not been
observed yet experimentally. This is probably the main reason for
such interest to it, and therefore search for a realistic scheme for
experimental observation of the DCE is an actual problem of modern
quantum electrodynamics.

A nonstationary cavity can be realized by two possible ways: (i) a
cavity with moving boundaries, or (ii) a cavity with fixed shape but
with {\it varying} boundary conditions. In case (i) a significant
number of photons could be created if the velocity of boundaries is
close to the speed of light. The corresponding experimental
realization would be obviously a difficult technical problem.
Alternatively, it was suggested to utilize the vibrating cavity and
accumulate the effect by tuning the frequency of small mechanical
oscillations of the boundaries (which could be induced, e.g., by a
sound wave \cite{Acoust}, or due to piezoelectric effect
\cite{Piezo}) in resonance with an eigenfrequency of the cavity
\cite{DManko,JR,Law,Dod,KA}, see also the recent review \cite{rev}.
In this realization, an extremely high accuracy of frequency tuning
must be provided.

Owing to amazing progress in modern laser technology and
semiconductor electronics, the variant (ii) could be considered as a
more realistic one \cite{Yab,Loz1,daccor}. The optical properties of
a semiconductor film located on a dielectric base can be changed,
e.g., by means of electron-hole plasma creation by a strong
femtosecond laser pulse \cite{Loz1}, or due to injection of carriers
by a powerful electric pulse \cite{FNL}. In both cases, variation of
optical properties of the boundary can be fast enough and well
controlled.

However, theoretical investigation of the first variant is simpler
from mathematical point of view, at least in the framework of a
one-dimensional model. Sometimes in the framework of the first
approach exact solutions to the problem can be found for some
special laws of motion. Besides, one may hope that very fast
displacement of a boundary acts like "almost instant" creation of a
new boundary due to changing of optical properties of the medium
inside the cavity. Therefore we will consider a cavity with moving
boundaries in this paper, and will be especially interested in the
case of ultrarelativistic motion.

We will present a new exact solution for the DCE problem in a
one-dimensional ideal cavity uniformly contracting during the time
interval $0<t<T$ and otherwise stationary. A complete set of
solutions for a classical field inside a one- or even
three-dimensional cavity  at the time of uniform contracting was
known for years, see Ref.~\cite{rev}. It was used, for example, in
Ref.~\cite{Bordag1,Bordag2}, for calculation of the Casimir force
acting between the boundaries of a relativistically squeezing or
expanding cavity. In contrast, our solution for a one-dimensional
quantum field in a uniformly contracting cavity seems to be unknown
in literature, and allows one to study the DCE for this particular
case non-perturbativly. We have discovered a new and unexpected
effect of periodic dependence of the number of created particles on
some variable $\chi_f$ determined by the time of contracting $T$ and
velocity of the boundary $v$. It is explained by quantum
interference of the events of particle creation which are taking
place only at the moments of acceleration ($t=0$) and deceleration
($t=T$) of a boundary, while stable particle states exist (and thus
no particles are created) at the time of contracting.

We review the Hamiltonian approach in Section \ref{Hamiltonian
preliminaries}. Consideration of a field quantization procedure in a
uniformly contracting cavity is presented in Section
\ref{quasiparticles}. Section \ref{particle creation} is devoted to
calculation of the energy spectrum of particles created in the
cavity. The periodic dependence of all measurable quantities on
$\chi_f$ is indicated, and the optimal conditions for the DCE are
derived. The discussion of the results and conclusions are given in
Section \ref{discussion}. We use natural units $\hbar=c=1$
throughout the paper.

\section{Hamiltonian and other preliminaries}
\label{Hamiltonian preliminaries}

We will use the Hamiltonian approach which in application to DCE was
apparently first formulated in Ref. \cite{RT}, see also
\cite{Law,rev}. In this section we begin with derivation of
Hamiltonian for a one-dimensional problem based on the well known
properties of time-dependent canonical transformations in classical
Hamiltonian mechanics (see, e.g., \cite{LLI}). Our approach clears
up the presence of extra terms in the Law Hamiltonian and is
equivalent to the method developed in the Law's paper \cite{Law}, at
least, for quantum systems with quadratic Hamiltonian.

Consider a massless scalar quantum field $\Phi(x,t)$ in a
nonstationary one-dimensional cavity formed by two mirrors. One of
the mirrors is fixed at the point $x=0$, while the position of the
other depends on time $x=l(t)$. For definiteness, we impose the
following boundary conditions at the mirrors:
$\Phi(0,t)=\Phi(l(t),t)=0$.

The Hamiltonian for the field reads
\begin{equation}\label{Hamiltonian} H=\frac12 \int\limits_0^{l(t)}
\left\{\left(\frac{\partial \Phi(x,t)}{\partial
t}\right)^2+\left(\frac{\partial \Phi(x,t)}{\partial
x}\right)^2\right\}\, dx.\end{equation} The field $\Phi(x,t)$ and
the canonical momentum $\Pi(x,t)=\partial\Phi(x,t)/\partial t$ are
assumed to obey the equal time commutation relations
$[\Phi(x,t),\Pi(x',t)]=i\delta(x-x')$, and hence the Heisenberg
equations acquire the form
\begin{equation}\label{Hamilt
Eqs}\dot\Phi(x,t)=i[H,\Phi(x,t)]=\Pi(x,t),\quad
\dot\Pi(x,t)=i[H,\Pi(x,t)]=\frac{\partial^2\Phi(x,t)}{\partial
x^2}.\end{equation}

One can easily see that the Fourier coefficients
\begin{eqnarray}\label{Q_n} Q_n(t)=
\sqrt{\frac{2}{l(t)}}\int\limits_0^{l(t)}\Phi(x,t)\sin\left(\frac{\pi
n x}{l(t)}\right)\,dx,\\ \label{P_n} P_n(t)=
\sqrt{\frac{2}{l(t)}}\int\limits_0^{l(t)}\Pi(x,t)\sin\left(\frac{\pi
n x}{l(t)}\right)\,dx,\end{eqnarray} satisfy the standard
commutation relations $[Q_n,P_{n'}]=i\delta_{nn'}$ and
$[Q_n,Q_{n'}]=[P_n,P_{n'}]=0$. Introducing the "instantaneous
eigenfrequencies" $\omega_n(t)=\pi n/l(t)$, and the "instantaneous
destruction and creation" operators $a_n$, $a_n^\dagger$
\begin{equation}\label{a_nac_n} a_n(t)=\frac{\omega_n(t)
Q_n(t)+iP_n(t)}{\sqrt{2\omega_n(t)}},\quad
a_n^\dagger(t)=\frac{\omega_n(t)
Q_n(t)-iP_n(t)}{\sqrt{2\omega_n(t)}},\end{equation} which satisfy
the canonical commutation relations
$[a_n,a_{n'}^\dagger]=\delta_{nn'}$,
$[a_n,a_{n'}]=[a_n^\dagger,a_{n'}^\dagger]=0$, one can write the
expansions for the field and canonical momentum in the form
\begin{equation}\label{a->Phi,Pi}\begin{array}{c}\displaystyle
\Phi(x,t)=\sqrt{\frac{2}{l(t)}}\sum\limits_{n}\sin\left(\frac{\pi
n x}{l(t)}\right)\frac{a_{n}(t)+a_{n}^\dagger(t)}{\sqrt{2\omega_{n}(t)}},\\
\\ \displaystyle
\Pi(x,t)=\sqrt{\frac{2}{l(t)}}\sum\limits_{n}\sin\left(\frac{\pi n
x}{l(t)}\right)\left(-i\sqrt{\frac{\omega_{n}(t)}{2}}\right)
\left\{a_{n}(t)-a_{n}^\dagger(t)\right\}.\end{array}\end{equation}
Since operators $a_n$, $ia_n^\dagger$ and $Q_n$, $P_n$ satisfy the
same commutation relations, Eqs.~(\ref{a->Phi,Pi}) can be considered
as a canonical transformation. One can easily check that this
transformation can be defined with a generating functional
$F[a,\Phi]$ as
\begin{equation}\label{canon_tranf}a_n^\dagger=-i\frac{\partial
F[a,\Phi]}{\partial a_n},\quad \Pi(x,t)=-\frac{\delta
F[a,\Phi]}{\delta\Phi(x,t)},\end{equation} \begin{equation}\label{F}
F[a,\Phi]=-\frac{i}{2}\sum\limits_n\left\{\omega_n(t)
Q_n^2[\Phi,t]+a_n^2-2\sqrt{2\omega_n(t)}Q_n[\Phi,t]
a_n\right\}.\end{equation} The order ambiguity in the last term of
Eq.~(\ref{F}) can give rise only to an additive indefinite c-number
constant, and thus is insignificant.

For the stationary case ($l(t)={\rm const}$) the generating
functional does not depend on time explicitly, so that the
Hamiltonian remains unchanged under the canonical transformation
(\ref{canon_tranf}),(\ref{F}). In terms of new variables it acquires
the form $H=\sum_n \omega_n a_n^\dagger a_n$. If the cavity is
nonstationary, and hence the generating functional explicitly
depends on time, the Hamiltonian changes, compare \cite{LLI}. The
operators $a_n$ satisfy the equations $\dot a_n=i[H',a_n]$, where
the new Hamiltonian is $H'=H-\partial F/\partial t$.

For the case of a uniformly contracting cavity
\begin{equation}\label{l(t)}l(t)=\left\{\begin{array}{ll}\displaystyle l_i,
&\displaystyle t<0,\\
\displaystyle l_i-vt,&\displaystyle 0<t<T,\\ \displaystyle
l_f,&\displaystyle t>T,\end{array}\right.\end{equation} where
$T=(l_f-l_i)/v$ is duration of contracting and $v>0$ is the velocity
of the moving boundary, we obtain at $0<t<T$
\begin{equation}\label{Lambda} H'=\pi
\Lambda/l(t),\quad \Lambda=\sum\limits_n \left\{n{a}_n^{\dagger}
a_n- i\frac{v}{4\pi}(a_n^2-{a_n^\dagger}^2)\right\}
 - \frac{iv}{\pi} \sum\limits_{n\ne n'}
(-1)^{n+n'}\sqrt{nn'}\times\left\{ \frac{(a_n
a_{n'}-{a}_{n'}^{\dagger}{a}_n^{\dagger})}{2(n+n')}+\frac{{a}_n^{\dagger}
a_{n'}}{n'-n}\right\}.\end{equation} The same result follows from
the general Law formula (2.21) of Ref.~\cite{Law} for the
"effective" Hamiltonian which was constructed to reduce the
equations of motion for operators $Q_n$, $P_n$ to the Heisenberg
form. This proves that both approaches are equivalent at least for
the case when the Hamiltonian is quadratic in structure.

The additional terms in (\ref{Lambda}) are responsible for
annihilation and creation of pairs and scattering of particles from
one mode to another in the contracting cavity. The corresponding
"coupling constants" are proportional to the velocity of the moving
boundary and are small in non-relativistic case. One can see that
the terms of the Hamiltonian (\ref{Lambda}) responsible for
interaction between modes are of the same order as those describing
particle production in a separate mode. Therefore strong interaction
between field modes is a built-in internal feature of DCE
\cite{Pomeran}.

We see from Eq.~(\ref{Lambda}) that the Hamiltonian $H'$ has a
remarkable property. Its time dependence is determined exclusively
by a c-number factor $\pi/l(t)$. Hence, the operator $\Lambda$ does
not depend on time explicitly and is an integral of motion.
Therefore, being quadratic in canonical variables $a_n$,
$a_n^\dagger$, it admits diagonalization by a time independent
linear canonical transformation (Bogolubov transformation) from
$a_n$, $a_n^\dagger$ to some new canonical variables $\tilde{b}_n$,
$\tilde{b}_n^\dagger$. In terms of these new variables we would have
\begin{equation}\label{H'}\Lambda=\sum_n\lambda_n\tilde{b}_n^
\dagger\tilde{b}_n\,,\quad
H'=\sum_n(\pi\lambda_n/l(t))\tilde{b}_n^\dagger\tilde{b}_n.
\end{equation}

Hereinafter it is convenient to introduce an auxiliary dimensionless
time variable $\chi=\pi\int_0^t dt/l(t)$, which varies from
$-\infty$ to $+\infty$. At $0<t<T,\;\chi=-(\pi/v)\ln(1-vt/l_i)$. In
terms of this variable the Heisenberg equations of motion acquire
the form
\begin{equation}\label{EffHamiltEqs}
da_n/d\chi=i[\Lambda,a_n],\quad 0<t<T,\end{equation} so that the
operator $\Lambda$ can be considered as the generator of
translations in "time" $\chi$. The time moment $t=0$ corresponds to
$\chi_i=0$, whereas $t=T$ to $\chi_f=(\pi/v)\ln(1/\rho)$, where the
dimensionless parameter $\rho=l_f/l_i<1$ is the squeeze rate for the
cavity. The value $\chi=\infty$ corresponds to the moment $t=l_i/v$
of the cavity collapse.

\section{Diagonalization of the Hamiltonian}\label{quasiparticles}

We will use the so-called Milne reference frame, see, e.g., Ref.
\cite{BD}, to diagonalize operator $\Lambda$. The corresponding
spatial $\xi$ and time $\tau$ Milne coordinates are defined by
\begin{equation}\label{Milne_Coords} t=\frac{l_i}{v}-\tau\cosh{\xi},\quad
x=\tau\sinh{\xi},\end{equation} The point is that the cavity remains
stationary in Milne reference frame. Indeed, the spacetime region
$0<x<l_i-vt$ is mapped conformally by transformation
(\ref{Milne_Coords}) onto the strip $0<\xi<(1/2)\ln{d}$, $\tau>0$,
where $d=(1+v)/(1-v)$ coincides with the Doppler factor for
reflection from a mirror moving with velocity $v$. This means that
the world line for the right boundary is given by
$\xi=(1/2)\ln{d}=const$.

The wave equation in terms of Milne coordinates reads:
\begin{equation}\label{Milne_WaveEq}\frac1{\tau}\frac{\partial}{\partial\tau}
\left(\tau\frac{\partial\Phi}{\partial\tau}\right)-
\frac1{\tau^2}\frac{\partial^2\Phi}{\partial\xi^2}=0.\end{equation}
This equation can be easily solved by separation of variables.
Taking into account boundary conditions (which are stationary in the
Milne reference frame) we obtain a complete set of modes
\begin{equation}\label{Milne_Modes} \Psi_n(\xi,\tau)=\frac1{\sqrt{\pi
n}} \left(\frac{v\tau}{l_i}\right)^{2\pi i
n/\ln{d}}\sin\left(\frac{2\pi n\xi}{\ln{d}}\right),\quad
n=1,2,\ldots\;,\end{equation} which are normalized by the relation
$$-i\int\limits_0^{\frac12\ln{d}} d\xi\,\tau
\left[\Psi_n^*\frac{\stackrel{\leftrightarrow}{\partial}}
{\partial\tau}\Psi_{n'}\right]=\delta_{nn'}.$$ Note that the unusual
sign in the LHS of this equation arises due to opposite directions
of physical time $t$ and Milne time $\tau$. In terms of the original
variables $x$, $t$ these modes acquire the form
\begin{equation}\label{Milne_ModesXT}
\Psi_n(x,t)=\frac{-i}{2\sqrt{\pi
n}}\left\{\left(1-\frac{v(t-x)}{l_i}\right)^{\frac{2\pi i
n}{\ln{d}}}-\left(1-\frac{v(t+x)}{l_i}\right)^{\frac{2\pi i
n}{\ln{d}}}\right\},\end{equation} and constitute a complete set of
solutions for the wave equation in a uniformly contracting cavity
orthonormalized by the usual Klein-Gordon scalar product
$$i\int\limits_0^{l(t)} dx\,
\left[\Psi_n^*\frac{\stackrel{\leftrightarrow}{\partial}} {\partial
t}\Psi_{n'}\right]=\delta_{nn'}.$$ Hence, the general solution to
the Heisenberg equations (\ref{Hamilt Eqs}) can be represented in
the form
\begin{equation}\label{VirtModeQuant}\Phi(x,t)=\sum\limits_n
\left\{b_n\Psi_n(x,t)+
b_n^\dagger\Psi_n^*(x,t)\right\},\end{equation} where operators
$b_n$, $b_n^\dagger$ satisfy the standard Bose-Einstein commutation
relations and {\it are independent of time} since modes
(\ref{Milne_ModesXT}) satisfy the wave equation exactly.
Specifically, this means that the Hamiltonian which governs behavior
of the variables $b_n$ and $b_n^\dagger$ is {\it equal to zero} with
an accuracy to a $c$-number contribution.

Let us now express operators $b_n$ in terms of operators $a_n$,
$a_n^\dagger$. Due to completeness and orthonormality of the set
(\ref{Milne_ModesXT}), we have:
\begin{equation}\label{Phi,Pi->b}
b_n=i\int\limits_0^{l(t)}\left\{\Psi_n^*(x,t)\Pi(x,t)-
\frac{\partial\Psi_n^*(x,t)}{\partial
t}\Phi(x,t)\right\}\,dx.\end{equation} Using expansions
(\ref{a->Phi,Pi}) we reduce RHS of Eq.~(\ref{Phi,Pi->b}) to the form
$$b_n=\sum\limits_{n'}\left\{A_{nn'}(t)a_{n'}(t)+B_{nn'}(t)a_{n'}^\dagger(t)\right\},$$
where
\begin{equation}\label{ABcoeffs}\left.\begin{array}{c}\displaystyle
A_{nn'}(t)\\ \displaystyle
B_{nn'}(t)\end{array}\right\}=\frac{-i}{\sqrt{\pi
n'}}\int\limits_0^{l(t)}dx\,\sin\left(\frac{\pi
n'x}{l(t)}\right)\left\{\frac{\partial\Psi_n^*(x,t)}{\partial t}\pm
i\omega_{n'}\Psi_n^*(x,t)\right\}.\end{equation} Now, using explicit
expressions (\ref{Milne_ModesXT}) we can express time derivatives in
Eq.~(\ref{ABcoeffs}) in terms of derivatives with respect to $x$ and
perform integration by parts. Introducing then a scaling variable
$y=x/l(t)$ we finally obtain:
\begin{equation}\label{AB->alpha,beta} A_{nn'}(t)=e^{2iv
n\chi(t)/\ln{d}}\alpha_{nn'},\quad B_{nn'}(t)=e^{2iv
n\chi(t)/\ln{d}}\beta_{nn'},\end{equation} where coefficients
\begin{equation}\label{alpha_beta}\left.\begin{array}{c}\displaystyle
\alpha_{nn'}\\ \displaystyle \beta_{nn'}\end{array}\right\}
=\frac12\,\sqrt{\frac{n'}n}\int\limits_{-1}^1 dy\,(1-vy)^{-2\pi i
n/\ln{d}}\,e^{\mp i\pi n'y},\end{equation} are already independent
of time $t$, or auxiliary variable $\chi$.

Thus, we have
\begin{equation}\label{b_tilde b}b_n=e^{2iv n\chi(t)/\ln{d}}\tilde{b}_n(t),
\end{equation}
where operators
\begin{equation}\label{tilde b}\tilde{b}_n(t)=\sum\limits_{n'}
\left\{\alpha_{nn'}a_{n'}(t)+\beta_{nn'}a_{n'}^\dagger(t)\right\},\end{equation}
are expressed in terms of operators $a_n$, $a_n^\dagger$ by a
Bogolubov transformation with coefficients, which {\it do not depend
on time explicitly}$\,$. Therefore the Hamiltonian which determines
time dependence of the operators $\tilde{b}_n$ remains unchanged
under transformation (\ref{tilde b}) and is given by
Eq.~(\ref{Lambda}). This result, together with Eq.~(\ref{b_tilde
b}), is sufficient to ascertain the structure of the Hamiltonian
$\Lambda$ in terms of the operators $\tilde{b}_n$.

Indeed, the Heisenberg equations of motion for operators
$\tilde{b}_n$ are given by
$d\tilde{b}_n/d\chi=i[\Lambda,\tilde{b}_n]$, compare
(\ref{EffHamiltEqs}). Taking into account the form (\ref{b_tilde b})
of time dependence of $\tilde{b}_n$, we have for the commutator
$[\Lambda,\tilde{b}_n]$:
$$[\Lambda,\tilde{b}_n]=-(2v/\ln d) n\tilde{b}_n\,.$$ It clearly follows
from (\ref{Lambda}) and (\ref{tilde b}) that $\Lambda$ is quadratic
in terms of operators $\tilde{b}_n$ also. Hence, we unambiguously
arrive to
\begin{equation}\label{Lambda_diag}\Lambda=\frac{2v}{\ln{d}}\sum\limits_n n
\tilde{b}_n^\dagger \tilde{b}_n\,.\end{equation}

Eqs.~(\ref{Lambda_diag}), (\ref{tilde b}) and (\ref{alpha_beta})
constitute an exact solution for quantum field in a uniformly
contracting cavity. This is one of the primary results of the
present paper. Its physical meaning consists in existence of a
notion of stable, non-interacting particles in a uniformly
contracting one-dimensional cavity. Since pair production or
mode-to-mode scattering are absent in this representation, the
number of these particles in every mode is fixed. Their energies,
however, depend on time due to collisions with the moving boundary,
compare Eq.~(\ref{H'}). It is worth noting that the exact energy
spectrum of these particles turns out to be equidistant.

It is important that the operators $\tilde{b}_n(t)$ for the case of
a stationary cavity coincide with $a_n(t)$. Indeed, proceeding to
the limit $v\rightarrow 0$ in Eq.~(\ref{alpha_beta}), one can easily
make certain that $\alpha_{n,n'}(0)=\delta_{nn'},\,
\beta_{n,n'}(0)=0$. We may conclude that operators
$\tilde{b}_n(t),\,\tilde{b}^{\dagger}_n(t)$ defined by
Eqs.~(\ref{tilde b}),(\ref{alpha_beta}) represent a natural
generalization for particle destruction and creation operators valid
for nonstationary cavities with uniformly moving boundaries.

Let us now consider the Bogolubov transformation (\ref{tilde b})
with coefficients $\alpha_{n,n'}(v),\,\beta_{n,n'}(v)$ defined by
Eq.~(\ref{alpha_beta}) for $v(t)$ being an arbitrary function of
time. Since coefficients of the transformation (\ref{tilde b}) now
depend on time, the operators $\tilde{b}_n(t)$ are ruled by a new
Hamiltonian $\widetilde{H}$, so that the Heisenberg equations of
motion acquire the form
$i\dot{\widetilde{b}}_n=[\widetilde{H},\widetilde{b}_n]$. Using
Eqs.~(\ref{tilde b}) and equations of motion for operators $a_n$, we
obtain for the time derivative of $\tilde{b}_n(t)$:
$$\begin{array}{l}\displaystyle
\dot{\widetilde{b}}_n=
\sum\limits_{n'}\left(\dot\alpha_{nn'}a_{n'}+\dot\beta_{nn'}a_{n'}^\dagger+
\alpha_{nn'}\dot a_{n'}+\beta_{nn'}\dot a_{n'}^\dagger\right)=\\
\\ \displaystyle
=\sum\limits_{n'}\left(\dot\alpha_{nn'}a_{n'}+\dot\beta_{nn'}a_{n'}^\dagger+
\alpha_{nn'} i[H',a_{n'}]+\beta_{nn'} i
[H',a_{n'}^\dagger]\right)=\sum\limits_{n'}\left(\dot\alpha_{nn'}a_{n'}+\dot\beta_{nn'}a_{n'}^\dagger\right)+
i[H'\widetilde{b}_n].\end{array}$$ We know already the structure of
$H'$ in terms of $\widetilde{b}_n$, see Eqs.~(\ref{Lambda}) and
(\ref{Lambda_diag}). For the difference $\widetilde{H}-H'\equiv{\cal
V}$ we have:
$$
i[{\cal V},\widetilde{b}_n]=
\sum\limits_{n'}\left(\dot\alpha_{nn'}a_{n'}+\dot\beta_{nn'}a_{n'}^\dagger\right).$$
Using the well known unitarity relations for Bogolubov coefficients
\begin{equation}\label{unitarity}
\sum\limits_{n''}\left(\alpha_{nn''}\beta_{n'n''}-
\alpha_{n'n''}\beta_{nn''}\right)=0,\quad
\sum\limits_{n''}\left(\alpha_{nn''}\alpha_{n'n''}^*-
\beta_{nn''}\beta_{n'n''}^*\right)=\delta_{nn'},\end{equation} which
in our case (\ref{alpha_beta}) can be verified straightforwardly, we
get the inverse transformation
\begin{equation}\label{a_inv}
a_n=\sum\limits_{n'}\left(\alpha_{n'n}^*\widetilde{b}_{n'}-
\beta_{n'n}\widetilde{b}_{n'}^\dagger\right),\end{equation} and
hence,
\begin{equation}\label{com_VB}
i[{\cal V},\widetilde{b}_n]= \dot v\sum\limits_{n'}\left({\cal
A}_{nn'}\widetilde{b}_{n'}+{\cal
B}_{nn'}\widetilde{b}_{n'}^\dagger\right),\quad
\end{equation}
where
\begin{equation}\label{calA}
{\cal
A}_{nn'}=\sum\limits_{n''}\left(\frac{d\alpha_{nn''}}{dv}\alpha_{n'n''}^*-
\beta_{n'n''}^*\frac{d\beta_{nn''}}{dv}\right),\quad {\cal
B}_{nn'}=\sum\limits_{n''}\left(\frac{d\beta_{nn''}}{dv}\alpha_{n'n''}-
\beta_{n'n''}\frac{d\alpha_{nn''}}{dv}\right).
\end{equation}
Substituting representations (\ref{alpha_beta}) for Bogolubov
coefficients into Eqs.~(\ref{calA}) we can perform summations over
$n''$ in (\ref{calA}) using the well known relation
$$\sum\limits_{n=-\infty}^{\infty}ne^{i\pi xn}=-\frac{2i}{\pi}
\sum\limits_{k=-\infty}^{\infty}\delta'(x-2k).$$ After this, all
integrations become trivial and we obtain:
\begin{equation}\label{calA_final}\begin{array}{c}\displaystyle
\mathcal{A}_{nn'}=i\frac{(-1)^{n'-n}}{\pi}
\frac{\sqrt{nn'}\gamma^{2+2i\pi(n'-n)/\ln d}}{(n'-n)(n'-n+i\ln
d/2\pi)}\,,\quad n\neq n'\,,\qquad \mathcal{A}_{nn}=-\frac{2i\pi
n}{\ln^2d} \bigg[\frac{\ln d}{v}+2\gamma^2(\ln\gamma)-1\bigg],\\ \\
\displaystyle\mathcal{B}_{nn'}=i\frac{(-1)^{n'+n}}{\pi}
\frac{\sqrt{nn'}\gamma^{2-2i\pi(n'+n)/\ln d}}{(n'+n)(n'+n-i\ln
d/2\pi)}\,,
\end{array}\end{equation}
where $\gamma=(1-v^2)^{-1/2}$.

It follows from (\ref{calA_final}) that ${\cal A}_{nn'}^*=-{\cal
A}_{n'n}$, ${\cal B}_{nn'}={\cal B}_{n'n}$. Using these relations
and Eq.~(\ref{com_VB}) one can easily find
\begin{equation}\label{V_cal}
{\cal V}=i\,\dot v(t)\sum\limits_{n,n'}\left\{{\cal
A}_{nn'}(v(t))\widetilde{b}_{n}^\dagger\widetilde{b}_{n'}+\frac12\left({\cal
B}_{nn'}(v(t))\widetilde{b}_{n}^\dagger\widetilde{b}_{n'}^\dagger-{\cal
B}_{nn'}^*(v(t))\widetilde{b}_{n}\widetilde{b}_{n'}\right)\right\}.
\end{equation} And finally
\begin{equation}\label{H_tilde}
\widetilde{H}=\frac{2\pi v(t)}{l(t)\ln{d}}\sum\limits_n n
\widetilde{b}_n^\dagger \widetilde{b}_n+{\cal V}.
\end{equation}

The representation (\ref{H_tilde}) is very important because it
clearly indicates that (i) the Hamiltonian for a field in a
nonstationary cavity can be diagonalized if $\dot{v}=0$, and hence,
we can use the notion of particles in a uniformly contracting
cavity; and (ii) particle creation, as well as mode-to-mode
scattering, takes place only when the boundary moves with
acceleration. For our law of motion (\ref{l(t)}) these processes
arise only in the initial $t=0$ and final $t=T$ moments of time. The
fact that radiation processes are absent during the time when one
boundary of a cavity moves with a constant velocity was first
noticed by Fulling and Davis in Ref.~(\cite{FD}).

\section{Particle creation in a uniformly contracting
cavity}\label{particle creation}

We will now compute the number of particles created in a
nonstationary cavity with the right mirror moving according to
Eq.~(\ref{l(t)}). We will note first that approach based on the
Hamiltonian $\widetilde{H}$ is not convenient for this purpose. This
is because velocity of the mirror in our approximation changes
instantaneously, and thus $\widetilde{H}$ contains ill defined
products of $\delta$ and $\theta$-functions. Thereagainst, the
Hamiltonian $H'$ (\ref{Lambda}) is not that singular even in our
approximation (it contains only $\theta$-functions) and therefore
operators $a_n(t)$ are continuous functions of time. This point
plays the central role for our method of solution.

We start from the initial conditions
\begin{equation}\label{in_cond}a_n(\chi)\vert_{\chi=-0}=a_n^{in},
\end{equation}
where $a_n^{in}$ are initial destruction operators which define the
initial vacuum state by $a_n^{in}|0\rangle_{in}=0$. Due to
continuity of operators $a_n(\chi)$ ($a_n(-0)=a_n(+0)$), we have the
initial data for the operators $\tilde b_n(\chi)$,
\begin{equation}\label{tilde b_in}\tilde{b}_n(0)=\sum\limits_{n'}
\left\{\alpha_{nn'}a_{n'}^{in}+\beta_{nn'}a_{n'}^{in\,\dagger}\right\}.
\end{equation}
According to Eq.~(\ref{b_tilde b}) the operator $\widetilde{b}_n$
depends on $t$ (or $\chi$) by mean of phase rotation. Specifically,
at $t=T$, we have:
\begin{equation}\label{bin->bout}\tilde b_n(\chi_f)=e^{-2\pi i n
\Theta}\tilde{b}_n(0),\quad
\Theta=\frac{\ln(1/\rho)}{\ln{d}}\,.\end{equation} Here the quantity
$\Theta$ measures the phase obtained by a particle in the principle
mode ($n=1$) during the squeezing period of the cavity. As it
follows from Eq.~(\ref{bin->bout}), the phase factor is completely
determined by the fractional part of $\Theta$. For further
convenience, we will denote the fractional part of $\Theta$ by
$\vartheta$,
\begin{equation}\label{Q-def1}
\vartheta={\rm
Frac}\left\{\frac{\ln(1/\rho)}{\ln{d}}\right\}.\end{equation} The
range of variation of the quantity $\vartheta$ is $0\le\vartheta<1$
(the symbol ${\rm Frac}$ denotes the fractional part of a number).

The operators $a_n(\chi_f-0)$, which are the solutions for
Eqs.~(\ref{EffHamiltEqs}) with the initial conditions
(\ref{in_cond}), can be now found from Eq.~(\ref{a_inv}). Hence,
again taking into account the continuity of operators $a_n$, we
finally have for $a_n^{out}=\\a_n(\chi_f+0)$:
\begin{equation}\label{tilde a_out}a_n^{out}=\sum\limits_{n'}
\left\{\alpha_{n'n}^*\tilde{b}_{n'}(\chi_f)-
\beta_{n'n}\tilde{b}_{n'}^\dagger(\chi_f)\right\}. \end{equation} By
combining the three transformations (\ref{tilde b_in}),
(\ref{bin->bout}) and (\ref{tilde a_out}), we obtain:
\begin{equation}\label{Bogol_tranf}a_n^{out}=\sum\limits_{n'}
\left\{U_{nn'}a_{n'}^{in}+V_{nn'}a_{n'}^{in\dagger}\right\},
\end{equation}
where the coefficients of the Bogolubov transformation
(\ref{Bogol_tranf}) are given by
\begin{eqnarray}\label{U_coeff}
U_{nn'}=\sum\limits_{n''}\left\{e^{-2\pi i n''
\vartheta}\alpha_{n''n}^*\alpha_{n''n'}-e^{2\pi i n''
\vartheta}\beta_{n''n'}^*\beta_{n''n}\right\},\\
\label{V_coeff} V_{nn'}=\sum\limits_{n''}\left\{e^{-2\pi i n''
\vartheta}\alpha_{n''n}^*\beta_{n''n'}-e^{2\pi i n''
\vartheta}\alpha_{n''n'}^*\beta_{n''n}\right\}.\end{eqnarray}

Consider the expression for $V_{nn'}$. Using Eq.~(\ref{alpha_beta}),
we get
\begin{equation}\label{V1}
V_{nn'}(d,\vartheta)=\frac{\sqrt{nn'}}{4}\int\limits_{-1}^{1}
dy_1\int\limits_{-1}^{1} dy_2\sum\limits_{n''\ge
1}\frac1{n''}\left(\frac{1-vy_1}{1-vy_2}\right)^{2\pi i
n''/\ln{d}}\times\left\{e^{i\pi(ny_1+n'y_2)-2\pi i n''
\vartheta}-e^{i\pi(n'y_1+ny_2)+2\pi i n''
\vartheta}\right\}.\end{equation} After evaluation of the sum and
double integral in (\ref{V1}) (see Appendix for details) we obtain
\begin{equation}\label{V_final}
V_{nn'}=(-1)^{n+n'}\frac{i(d-1)\sqrt{nn'}}{2\pi(n
d^{\vartheta}+n')(n+n'd^{1-\vartheta})}\times\left\{e^{-2 \pi i
n(d^{\vartheta}-1)/(d-1)}-e^{-2 \pi i
n'(d^{1-\vartheta}-1)/(d-1)}\right\}.\end{equation} The
coefficient $U_{nn'}$ differs from $V_{nn'}$ by the only change
$n'\to -n'$ everywhere, except for the common factor $\sqrt{n'}$,
so that
\begin{equation}\label{U_final}
U_{nn'}=(-1)^{n+n'}\frac{i(d-1)\sqrt{nn'}}{2\pi(n
d^{\vartheta}-n')(n-n'd^{1-\vartheta})}\times\left\{e^{-2 \pi i
n(d^{\vartheta}-1)/(d-1)}-e^{2 \pi i
n'(d^{1-\vartheta}-1)/(d-1)}\right\}.\end{equation} Finally, the
energy (mode) distribution of created particles is given by
\begin{equation}\label{Nn} \bar{N}_n
(d,\vartheta)=\,_{in}\langle
{0}|a_n^{out\,\dagger}a_n^{out}|0\rangle _{in}=\sum\limits_{n'\ge
1}|V_{nn'}|^2=\frac{(d-1)^2n}{\pi^2}\, \sum\limits_{n'=1}^{\infty}\,
\frac{n'\,\sin^2\left[\frac{\pi(d^{\vartheta}-1)}{(d-1)} (n+n'
d^{1-\vartheta})\right]}{(n d^{\vartheta}+n')^2\,
(n+n'd^{1-\vartheta})^2}.\end{equation}

One can see that the number of particles (\ref{Nn}) created in each
mode depends on two parameters. One of them is the Doppler factor
$d$, which can be expressed either in terms of velocity of the
moving boundary, $d=(1+v)/(1-v)$, or in terms of the corresponding
Lorentz factor $\gamma=(1-v^2)^{-1/2}$,
$d=\left(\gamma+\sqrt{\gamma^2-1}\right)^2$. The second parameter
$\vartheta$ is determined by $d$ and a dimensionless squeeze rate of
the cavity $\rho=l_f/l_i$ which, in turn, can be expressed in terms
of the duration of squeezing $T$ as $\rho=1-vT/l_i$. Note that
acceleration of the moving boundary,
$$
W=-v\delta(t)+v\delta(t-T),
$$
is determined by the same set of parameters. It is natural, since
particles can be created by an accelerating boundary only.

An interesting feature of the obtained solution is that the number
of created particles is a periodic function of the parameter
$\log\rho$ with the period $\log d$. This can be traced already from
the expansions (\ref{U_coeff}), (\ref{V_coeff}), and definition
(\ref{Q-def1}) of the parameter $\vartheta$. Surprisingly, the
number of created particles completely vanishes if $\vartheta=0$,
see Eq.~(\ref{Nn}). It is worth noting that $\bar{N}_n$ tends to
zero also if $\vartheta\to 1$, so that the energy spectrum of
created particles is a continuous function of parameter $\rho$.
\begin{figure}[htb]
\begin{center}\parbox{7cm}{\mbox{\epsfxsize=180pt
\epsffile{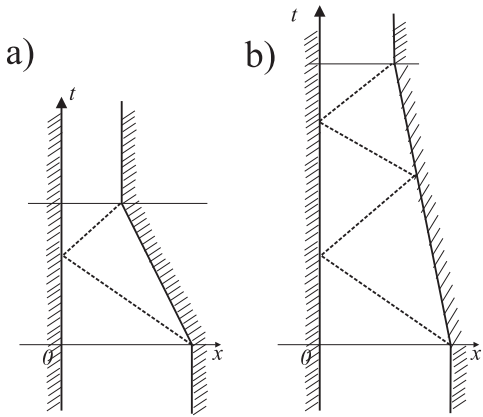}}}\end{center} \caption{\small A graphical
illustration for null values of the number of created particles at
a) $\rho=1/d$, b) $\rho=1/d^2$. The dashed lines represent world
lines of a light ray, emitted by the moving boundary at $t=0$.}
\label{PCzero}
\end{figure}
According to Eq.~(\ref{Q-def1}), $\vartheta$ turns into zero, and
hence the number of created particles vanishes, if $\rho=1/d^k$,
where $k=0,1,2,\ldots$. The value $k=0$ corresponds to a stationary
case, while the others can be explained by destructive quantum
interference between events of particle creation at the moments of
acceleration ($t=0$) and deceleration ($t=T$) of the boundary. As it
is illustrated by Fig.~\ref{PCzero}, the values of the squeeze rate
$\rho=1/d^k$ exactly correspond to the cases when a light ray,
emitted by the right boundary at $t=0$, returns to it at $t=T$ after
$k-1$ successive reflections and therefore can be absorbed at the
moment of deceleration. Analogous arguments were employed by the
authors of Ref.~(\cite{FD}) in discussion of the effect of vanishing
of the renormalized energy flux in a uniformly expanding cavity,
though only the case $k=1$ was considered in that work. It is worth
mentioning that at $\rho=1/d^k$ every level of the initial energy
spectrum of the field inside the cavity converts in time interval
$T$ to the value $\omega_n^f=\pi n/l_i\rho=(\pi n/l_i)d^k$
coinciding with the frequency acquired by a particle after $k-1$
successive reflections from the moving boundary, if this particle
was emitted by the boundary moving with velocity $v$ at $t=0$.

\begin{figure}[ht]
\begin{tabular}{ccc}
\epsfxsize8cm\epsffile{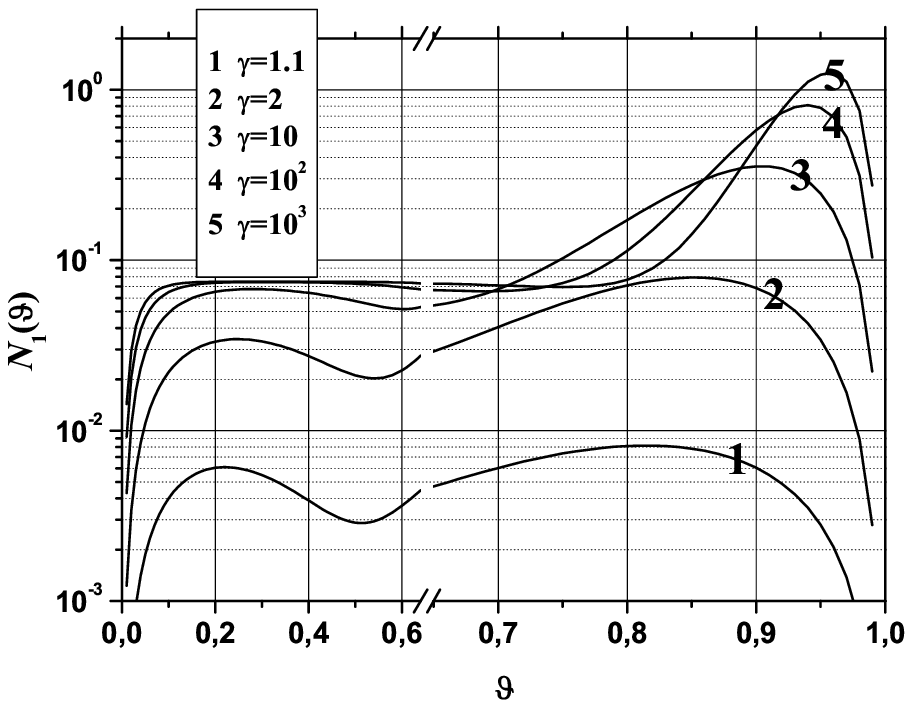} &
\epsfxsize8cm\epsffile{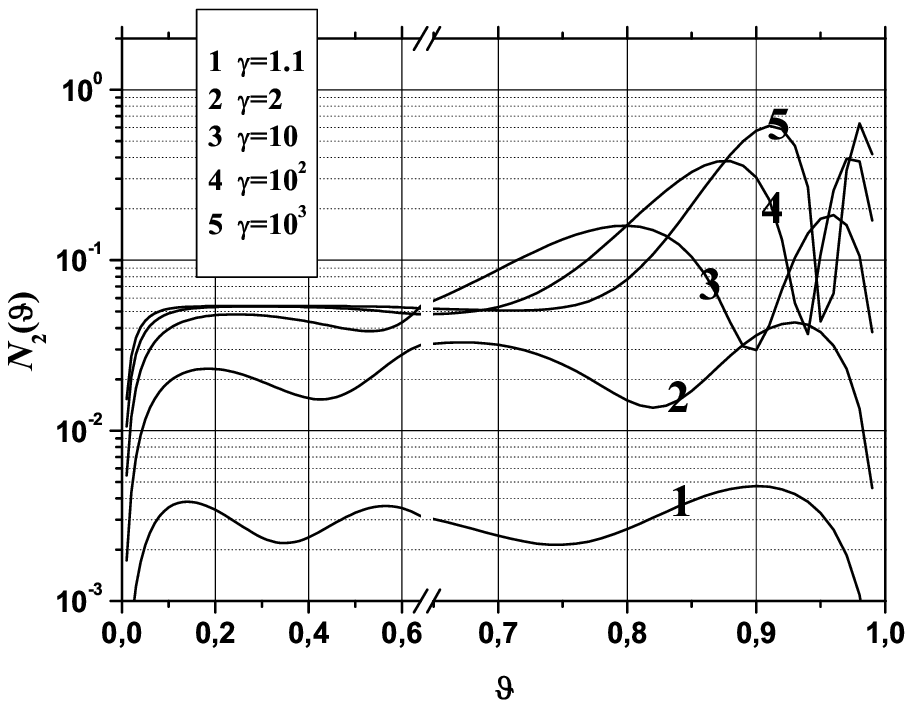} &
\\ a) & b) &
\end{tabular}
\caption{\small The average number of particles created in a) the
principle mode ($n=1$), and b) the first excited mode ($n=2$) as a
function of $\vartheta$ for different values of the moving boundary
$\gamma$-factor.} \label{N(Q)plot}
\end{figure}

The average number of particles, created in the principle ($n=1$)
and in the first excited ($n=2$) modes as functions of $\vartheta$
are shown in Fig.~\ref{N(Q)plot} for several
values of the moving boundary $\gamma$-factor. The same plots after
the appropriate scaling represent the first period of $\bar{N}_n$
dependence on $\log(1/\rho)$.

One can see from Fig.~\ref{N(Q)plot} that at sufficiently large
$\gamma$ plots for both $\bar{N}_1$ and $\bar{N}_2$ have a plateau
at not too small $\vartheta$, and a strongly pronounced maxima (one
for $n=1$ and two for $n=2$) at $\vartheta$ close to unity. It is
seen also that the altitudes of the plateaus are growing with
$\gamma$ remaining however rather small. To estimate the maximum
possible altitude of a plateau we will note that in the limiting
case $\gamma\to\infty$ ($d\to\infty$) at a fixed $\rho$ the
instantaneous approximation considered in Ref.~\cite{FNL} becomes
applicable. Indeed, one can easily make certain that Eq.~(\ref{Nn})
reduces to Eq.~(6) of Ref.~\cite{FNL} under condition $d\gg
n(1-\rho)/\rho$ \footnote{Note, that this condition of applicability
of instantaneous approximation was obtained in Ref.~\cite{FNL} on
the basis of merely qualitative arguments.}. For our estimation we
need only high-energy tail of energy distribution in instantaneous
approximation (see Eq.~(10) of Ref.~\cite{FNL}) where we should put
$\rho=0$ since not too small values of $\vartheta$ at $d\gg 1$
correspond to small $\rho$. Thus, for maximum values of $\bar{N}_n$
in the plateau region we have
\begin{equation}\label{plateau}\bar
N_n^{(pl)}\approx\frac1{2\pi^2n}\left[\ln(2\pi
n)+C-1\right],\end{equation} where $C=0.5572$ is the Euler constant.
For $n=1,2$ we have $\bar N_1^{(pl)}\approx 7.17\cdot 10^{-2}$ and
$\bar N_2^{(pl)}\approx 5.34\cdot 10^{-2}$, respectively. So, in the
plateau region the average number of created particles is small even
for ultrarelativistic velocities of the moving boundary.

To estimate positions of the maxima and the peak values of
$\bar{N}_n$ we should evaluate the sum in Eq.~(\ref{Nn}) under the
conditions $d\gg 1$, $1-\vartheta\ll 1$. It is clear that the major
contribution to the sum is given by $n'_{eff}\lesssim
nd^{\vartheta}$. Thus we have:
$$\bar N_n\approx\frac{n}{\pi^2}\, \sum\limits_{n'=1}^{\infty}\,
\frac{n'\,\sin^2\left[\pi(n d^{\vartheta-1}-n'
d^{-{\vartheta}})\right]}{(n+n' d^{-\vartheta})^2\, [n'+n
d^{\vartheta-1}]^2}\approx \frac1{\pi^2n}\,\sin^2\left(\pi n
d^{\vartheta-1}\right)\sum\limits_{n'=1}^{n d^{\vartheta}}\,
\frac{n'}{(n'+n d^{\vartheta-1})^2}\,.$$ Since $nd^{\vartheta}\gg
1$, we can change summation over $n'$ by integration. Finally, after
evaluation of the integral we obtain in the leading logarithmic
approximation:
\begin{equation}
\label{Nnexuni2}\displaystyle \bar
N_n\approx\frac1{\pi^2n}\sin^2\left(\pi n
d^{\vartheta-1}\right)\ln(n d).\end{equation}

We see from Eq.~(\ref{Nnexuni2}) that positions of the maxima are
determined by the relations
\begin{equation}\label{Res0}\pi n d^{\vartheta-1}=\pi (j+1/2),
\quad j=0,1,\ldots (n-1).
\end{equation}
This is an approximate condition but it becomes exact in the
ultrarelativistic limit. For example, at $\gamma=2,10,10^2,10^3$ for
the single maximum position at $n=1$ we have respectively
$\vartheta_{max}=0.74,0.88,0.94,0.95$ from Eq.~(\ref{Res0}), and
$\vartheta_{max}=0.85,0.90,0.94,0.96$ as the result of computation
according to exact formula (\ref{Nn}).

The condition (\ref{Res0}) can be reformulated in terms of
wavelengths as
\begin{equation}\label{Res} (2j+1)\lambda_n^f/2
=2l_i/d^{k+1},\end{equation} where $\lambda_n^f=2l_f/n$ is the
wavelength corresponding to the $n$-th harmonic in the final state.
The physical meaning of this condition can be clarified by the
following qualitative consideration.

\begin{figure}[htb]
\begin{center}\parbox{9cm}{\mbox{\epsfxsize=220pt
\epsffile{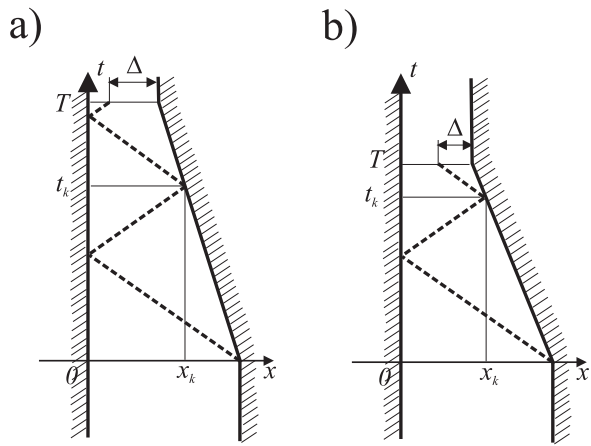}}}\end{center} \caption{\small World lines of
cavity boundaries (solid lines) and of the cluster of particles
(dashed line) created at $t=0$ for a) $T-t_k>x_k$, and b)
$T-t_k<x_k$. The number of collisions with the moving boundary
$k=1$.}\label{PCmax}
\end{figure}

We have shown, see Eqs.~(\ref{H_tilde}),(\ref{V_cal}), that
particles are created inside the cavity only at the moments of start
($t=0$) and stopping ($t=T$) of its boundaries. Physically, it is
clear that in the framework of the applied approximation
(instantaneous acceleration) the created particles constitute an
extremely narrow wave packet (cluster) which propagates then inside
the cavity without spreading successively colliding with its
boundaries, see Fig.~\ref{PCmax}. Such picture agrees with the
results of calculation of the energy momentum tensor performed by
Fulling and Davies in Ref.~(\cite{FD}).

Let us describe the clusters of particles created at the moments
$t=0$ and $t=T$ by classical fields $\Phi_1(x,t)$ and $\Phi_2(x,t)$
respectively. The function $\Phi_2(x,T)$ describes the second
cluster produced by the boundary stopping at the moment $t=T$. It is
located exactly at the point $x=l_f$. The first cluster at the same
moment of time is located at some distance $\Delta$ left to the
first one. The form of the cluster $\Phi_1$ is absolutely the same
and differs from $\Phi_2$ only by sign since the boundary
accelerations at two moments of particle creation were of opposite
sign. Thus, $\Phi_1(x,T)=-\Phi_2(x-\Delta,T)$. Expanding the fields
$\Phi_i$ in terms of stationary {\it out}-modes
$$\psi_{n}^{out}(x,t)=\frac{1}{\sqrt{\pi
n}}\sin\left(\frac{\pi n}{l_f}x\right)e^{-i\omega_{fn}t},\quad
\omega_{fn}=(\pi/l_f) n,$$ for the total field
$\Phi(x,T)=\Phi_1(x,T)+\Phi_2(x,T)$ we obtain
$$\Phi(x,T)=\sum_{n=1}^{\infty}\frac{2}{\sqrt{\pi n}}\Biggl(
c_{n}e^{-i\omega_{fn}T}+c_{n}^*e^{i\omega_{fn}T}
\Biggr)\cos\left(\frac{\pi
n}{l_f}(x-\frac{\Delta}{2})\right)\sin\left(\frac{\pi
n}{2l_f}\Delta\right).$$ Thus, we see that contribution of the
$n$-th mode to the field $\Phi(x,T)$ is absent if
$\,n\Delta/2l_f=j,\;j=0,1,2, ...$ (destructive interference of two
clusters), or it is maximal if
\begin{equation}\label{Delta}n\Delta/2l_f=j+1/2,\quad j=0,1,2, ...
\end{equation}
(constructive interference). In view of the restriction $\Delta\leq
l_f$, we have $2j+1\leq n$. Since particle creation and mode
interaction are absent in a stationary cavity, these conditions hold
at arbitrary moment of time $t>T$. Now, the question is whether it
is possible to match Eq.~(\ref{Delta}) with conditions (\ref{Res})
(as well as the corresponding conditions for minima).

Assume that the first cluster undergoes $k$ collisions with the
moving boundary during the period of contraction $T$. It can be
easily found that the last collision takes place at the point
$x_k=l_i/d^k$ at the moment of time $t_k=(l_i/v)(1-1/d^k)$. During
the period of time $T-t_k$ the cluster either collides with the
boundary at rest, if $T-t_k>x_k$ (Fig.~\ref{PCmax}a), or not, if
$T-t_k<x_k$ (Fig.~\ref{PCmax}b).
The quantity $\Delta$ for these
cases is given by the following relations respectively:
$$\Delta=l_f\pm[x_k-(T-t_k)].$$
Then Eq.~(\ref{Delta}) is equivalent to two relations
\begin{equation}\label{cond}
\frac{l_i}{d^k}\frac{1-v}{v}=\left\{\begin{array}{ll}\displaystyle
\frac{2l_f}{n}\Bigl(n\frac{1+v} {2v}-j-\frac{1}{2}\Bigr)\,,\quad
T-t_k>x_k\, \\ \\ \displaystyle\frac{2l_f}{n}\Bigl(j+\frac{1}{2}
+n\frac{1-v}{v}\Bigr)\,,\quad T-t_k<x_k\,
\end{array}\right.,
\end{equation}
where $j\leq (n-1)/2$. In the ultrarelativistic limit $1-v\ll 1$
($d=2/(1-v)$) the conditions (\ref{cond}) take on form of
Eq.~(\ref{Res}) with $0\leq j\leq n-1$. Our analysis shows that
physically the conditions (\ref{Res}) mean that maxima of $N_n$
appear in the cases when the first cluster is located at a distance
equal to an odd number of half-waves $\lambda_n^f/2$ from the point
of creation of the second cluster. It is worth emphasizing that
kinematics of our problem permits  implementation of this condition
only in the ultrarelativistic case. Therefore the less is the speed
of the boundary $v$, the less pronounced are the maxima and
$N_n\neq0$ in minima, compare Fig.~\ref{N(Q)plot}.

The values of $\widetilde{N}_n$ in the points of maxima are given by
Eq.~(\ref{Nnexuni2}) with logarithmic accuracy, and thus cannot
reproduce numerical results just as well as their positions.
However, Eq.~(\ref{Nnexuni2}) reproduces the main features of
spectra nearby the maxima qualitatively correctly. It is important
that the height of the maxima grows as $\log{d}$, or equivalently,
as $\log{\gamma}$ in the ultrarelativistic limit due to strong
interactions between the modes. Clearly, the effect of DCE
amplification arises due to constructive interference of events of
particle creation at $t=0$ and $t=T$. For example, if $\gamma=10^3$,
then for certain values of $\rho$ (or $T$) in the mean more than one
particle in the principle mode (or a little bit less than one
particle in the first excited mode) can be created.

\section{Discussion}\label{discussion}

An exact solution to the problem of dynamical Casimir effect in a
one-dimensional uniformly contracting cavity was found. This became
possible due to existence of stable non-interacting particle states
in such a cavity. As a consequence, we succeeded in diagonalization
of the Hamiltonian for a many-mode quantum field in a cavity with
uniformly moving boundaries (a generalization to the case with both
boundaries moving uniformly is straightforward, but rather
cumbersome). The developed method opens us possibilities to solve
exactly many other one-dimensional DCE problems for cavities with
boundaries moving along polygon-like world lines, including the
interesting case of vibrating boundaries. A new representation for
the field Hamiltonian was proposed in one-dimensional problem. It
was demonstrated explicitly that particle creation as well, as
mode-to-mode interaction is absent in a nonstationary cavity if its
boundaries are moving uniformly. However, the same approach can not
be applied to physically more realistic three-dimensional problem,
although a complete set of classical solutions generalizing
Eq.~(\ref{Milne_ModesXT}) is known for this case as well \cite{rev}.
This is due to breakdown of stability of particle states and
presence of interactions between them in three-dimensional problem.

It was shown that the number of particles, created in each mode,
depends periodically on the logarithm of the squeezing rate of the
cavity $\rho$, which is equal to the ratio of the final and initial
sizes of the cavity and is related to duration of contracting in a
rather simple way. This dependence is not monotonous. For the given
velocity of the moving boundary, there exists an infinite sequence
of values $\rho_k$ for which particles are not produced. On the
other hand, it is possible to specify an infinite set of
combinations of parameters, which correspond to an optimal regime of
particle production. Theoretically, at sufficiently large Lorentz
factor $\gamma$ for the moving boundary, the number of created
particles can be done arbitrary large. It is worth noting, however,
that the mentioned values of Lorentz factors are indeed very large.
For example, the value $\gamma\sim 10^3$ for production of a single
particle is necessary. We should also mention that the number of
created particles in the case of a cavity with ultrarelativistic
boundaries depends on $\gamma$ rather slowly (proportional to
$\log{\gamma}$).

The non-monotonous character of dependence of the number of created
particles on $\rho$ is also present in the framework of an
oversimplified "one-mode" model. However, correct estimations,
especially near the maxima of particle production rate, and even
correct qualitative understanding require taking into account
mode-to-mode interactions. This is true in principle at
consideration of any non-perturbative feature of DCE. Therefore, it
is very important to consider exactly soluble models. It should be
noticed, that the approximation of uniform motion of the boundaries,
separated by moments of instant infinite accelerations, does not
work well for high-frequency modes. For example, the energy spectrum
(\ref{Nn}) has an inadmissibly slowly decreasing tail ($\bar
N_n\propto 1/n$) leading to an infinite total number of created
particles. These problems can be resolved by smoothing instant
infinite accelerations, but the corresponding problem ceases be
exactly soluble.

\section*{Acknowledgments}

We thank M.I. Gozman, V.D. Mur, V.S. Popov and participants of the
Fourth D.N. Klyshko seminar for discussion of the results and
helpful remarks. This work was supported by the Russian Foundation
for Basic Research (grant 06-02-17370-a), by the Ministry of Science
and Education of Russian Federation (grant RNP 2.11.1972), and by
the Russian Federation President grants MK-2279.2005.2 and
NSh-320.2006.2.

\appendix
\section{Evaluation of the integral in Eq.~(\ref{V1})}
The coefficient $V_{nn'}$ (\ref{V1}) is represented by a difference
of two integrals. After the change of integration variables
$y_1\leftrightarrow y_2$ in the second of them, $V_{nn'}$ reduces to
the form:
\begin{equation}\label{V2} V_{nn'}=\frac{i\sqrt{nn'}}{2}\int\limits_{-1}^{1}
dy_1\int\limits_{-1}^{1} dy_2\, e^{i\pi(ny_1+n'y_2)}{\rm
Im}\,\sum\limits_{n''\ge 1}\frac{e^{2\pi i n''
Z(y_1,y_2)}}{n''},\end{equation} where
\begin{equation}\label{Z(y1,y2)}Z(y_1,y_2)=
\log_{d}\left(d^{-\vartheta}\frac{1-vy_1}{1-vy_2}\right).\end{equation}
For calculation of the sum in Eq.~(\ref{V2}) we will use formula
{\bf 1.441}.1 from \cite{RG} generalized for arbitrary values of $Z$
\begin{equation}\label{sum}
{\rm Im}\,\sum\limits_{n''\ge 1}\frac{e^{2\pi i n''
Z}}{n''}=\pi\left(\frac12-Z\right)+\pi\sum\limits_{k=-\infty}^{\infty}
k\theta(Z-k)\theta(k+1-Z)\,.
\end{equation}
Obviously, the first term in the RHS of (\ref{sum}) gives zero
contribution to the integral (\ref{V2}). Taking into account that
$-1-\vartheta\le Z(y_1,y_2)\le 1-\vartheta$ inside the integration
domain, and $0\le \vartheta<1$ by definition, we obtain
\begin{equation}\label{V3}\begin{array}{r}\displaystyle
V_{nn'}=-\frac{i\pi\sqrt{nn'}}{2}\int\limits_{-1}^{1}
dy_1\int\limits_{-1}^{1} dy_2\,
e^{i\pi(ny_1+n'y_2)}\left\{\theta(-Z)\theta(Z+1)
\right.\\\\
\left.\displaystyle
+2\theta(-Z-1)\theta(Z+2)\right\}.\end{array}\end{equation} It is
clear that the combination of $\theta$-functions
$$\theta(Z)+\theta(-Z)\theta(Z+1)+\theta(-Z-1)\theta(Z+2)$$
is equivalent to $1$ in the domain of integration, and thus
$$
\int\hspace{-0.2cm}\int\hspace{-0.1cm}d^2y\,
e^{i\pi(ny_1+n'y_2)}\left\{\theta(Z)+\theta(-Z)\theta(Z+1)+
\theta(-Z-1)\theta(Z+2)\right\}=0.$$ Therefore, we have:
\begin{equation}\label{V4}
V_{nn'}=\frac{i\pi\sqrt{nn'}}{2}\int\hspace{-0.2cm}\int \,d^2y\,
e^{i\pi(ny_1+n'y_2)}\left\{\theta(Z)-\theta(-Z-1)\right\}.
\end{equation}
The integral (\ref{V4}) can be easily reduced to the form
\begin{equation}\label{I}
V_{nn'}=\frac{i\pi\sqrt{nn'}}{2}\left\{\int\limits_{-1}^{a}
dy_1\int\limits_{b(y_1)}^1 dy_2\,-\int\limits_{a}^1
dy_1\int\limits_{-1}^{c(y_1)} dy_2\,\right\}
e^{i\pi(ny_1+n'y_2)}\,,\end{equation} where
$a=1-2(d^{\vartheta}-1)/(d-1)$,
$b(y_1)=(1-d^{-\vartheta})/v+d^{-\vartheta}y_1$, and
$c(y_1)=(1-d^{1-\vartheta})/v+d^{1-\vartheta}y_1$. After simple
calculations we now obtain the expression~(\ref{V_final}) for
$V_{n,n'}$.


\begin{references}
\bibitem{Moore}Moore G T 1970 \JMP {\bf 11} 2679.
\bibitem{Acoust}Dodonov V V 1995, \PL A {\bf 207} 126;
Dodonov V V, Klimov A B 1996, \PR A {\bf 53} 2664.
\bibitem{Piezo}Dodonov V V and Klimov A B 1996 \PR A {\bf 53}
2664.
\bibitem{DManko}Dodonov V V,  Man'ko V I  and Man'ko O V 1991
{\it Proceedings of Lebedev Physics Institute} vol 200 (Moscow:
Nauka) p 155.
\bibitem{JR}Jaekel M T and Reynaud S 1992.
{\it J. Phys. I (France)} {\bf 2} 149.
\bibitem{Law}Law C K 1994 \PR A {\bf 49} 433, 1995 \PR A {\bf 51} 2537.
\bibitem{Dod}Dodonov V V 1995 \PL A {\bf 207} 126.
\bibitem{KA}Klimov A B, Altuzar V 1997 \PL A {\bf 226} 41.
\bibitem{rev}Dodonov V V 2001 {\it Advances in Chemical Physics} vol
119 (John Wiley \& Sons, Inc) p~309.
\bibitem{Yab}Yablonovitch E 1989 \PRL {\bf 62} 1742.
\bibitem{Loz1}Lozovik Yu E, Tsvetus V G and Vinogradov E A 1995 \PS
{\bf 52} 284.
\bibitem{daccor}Dodonov  A V, Dodonov E V and Dodonov V V 2003, \PL A {\bf 317}
378.
\bibitem{FNL}Fedotov A, Narozhny N and Lozovik Yu 2005 \JOB {\bf 7}
S64.
\bibitem{Bordag1}M. Bordag, G. Petrov and D. Robaschik, Yad. Fiz. 39, 1315 (1984)
[Sov. J. Nucl. Phys. 39, 828 (1984)].
\bibitem{Bordag2} M. Bordag, F.-M. Dittes and D. Robaschik,
Yad. Fiz. 43, 1606 (1986) [Sov. J. Nucl. Phys. 43, 1034 (1986)].
\bibitem{RT}Razavy M, and Terning J 1985 \PR D {\bf 31} 307.
\bibitem{LLI}L. D. Landau and E. M. Lifshitz, {\it Mechanics}
(3rd Edition, Butterworth-Heinemann, 1981).
\bibitem{Pomeran}Yu.E. Lozovik, N.B. Narozhny and A.M.
Fedotov, Proceedings of the International Conference {\it I. Ya.
Pomeranchuk and Physics at the Turn of Centuries}, Edited by A.
Berkov, N. Narozhny and L. Okun, (World Scientific, Singapore,
2004), p.446.
\bibitem{BD}N.D. Birrell and P.C.W. Davies, {\it Quantum
Fields in Curved Space} (Cambridge University Press, Cambridge,
1982).
\bibitem{FD}Fulling S A and Davies P C W 1975, {\it Proc. R. Soc.
Lond.}A {\bf 348} 393.
\bibitem{RG} I.S. Gradshteyn, and I.M. Ryzhik, {\it Tables of
Integrals, Series, and Products} (Academic, New York, 1967).
\end{references}
\end{document}